\documentclass[a4paper,12pt]{article}
\usepackage[T2A]{fontenc}             
\usepackage[utf8x]{inputenc}          
\usepackage[english]{babel}   
\usepackage{babelbib}

\usepackage{amsmath,amsfonts,amssymb}
\usepackage{mathtext,nicefrac,cancel}
\usepackage{misccorr}
\usepackage[uline]{hhtensor}
\usepackage{xspace}
\usepackage{mathabx} 
\usepackage{epigraph}
\usepackage{setspace}  
\usepackage{xcolor}
\usepackage{url}

\usepackage{topcapt}
\usepackage[footnotesize,centerlast]{caption2}
\usepackage{graphicx}
\graphicspath{{images/}}

\usepackage[draft]{prelim2e}

\def\G{\ensuremath{\mathbf{G}}\xspace}

\usepackage{tikz}
\newcommand\centerofmass{\ensuremath{%
	\protect\tikz[scale=2.2,radius=0.4mm,black,line width=0.1mm]{
	\protect\draw (0,0) circle;
	\protect\fill[rotate=  0] (0,0) -- ++(0.4mm,0) arc (0:90:0.4mm);
	\protect\fill[rotate=180] (0,0) -- ++(0.4mm,0) arc (0:90:0.4mm);
	}}\xspace%
}

\begin{document}

\vspace*{-20mm}
\begin{center}
	\noindent
	\Large\textbf{On the effect of the central body small deformations
		on its satellite trajectory in\\ the problem of the two-body gravitational interaction}
\end{center}
\bigskip

\centerline{D.G. Kiryan and G.V. Kiryan}
\medskip

\begin{center}
	\begin{minipage}{110mm}
		\emph{\small Institute of Problems of Mechanical Engineering of RAS\\
		61 Bol'shoy pr. V.O., Saint-Petersburg 199178, Russia\\
		e-mail: \textit{diki.ipme@gmail.com}}
	\end{minipage}
\end{center}

\vspace*{-6mm}
%

\section*{}

\noindent
The problem of the two-body gravitational interaction has been solved numerically based on the classical mechanics principles. One of the bodies is a deformable three-axis ellipsoid (central body) and the other is a material point (satellite). The relationship of the angular discrepancy between the calculated and actual positions of the satellite pericenter with central body's gravity anomaly has been established.
\bigskip

\noindent
\textbf{Keywords:} gravitation, gravity prospecting, oblateness, barycenter


%

\section{Problem definition and justification}

\begin{figure}[h!]
	\centering
	\includegraphics[scale=1]{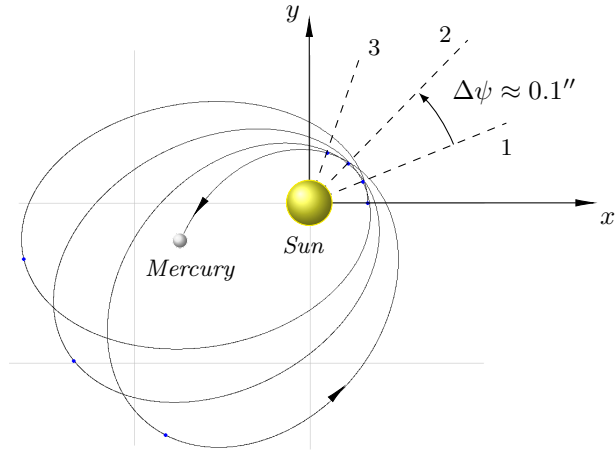}
	\caption{%
		''Anomalous'' shift of the Mercury's perihelion~$\Delta\psi$ during its synodic period. Digits $1,2,3$ mark the sequence of the Mercury's perihelia.
	}
	\label{fig:apsides7-fig-pericenter}
\end{figure}

\noindent
Analysis of the influence of nonsphericity of a massive central body upon its satellite trajectory is a well-known problem of satellite-based gravity prospecting. However, it has not been yet set out with respect to the Sun and Mercury, and, thus, is of a specific interest. The point of interest is to try, in terms of the classical mechanics, to discover the nature of the so-called ''anomalous'' shift of the Mercury's perihelion%
\footnote{%
	The minimal distance between the satellite and center of mass of the central body.
}
(Fig.~\ref{fig:apsides7-fig-pericenter}) during its synodic year%
\footnote{%
	The object's period of revolution about the Sun.
}
by angle~$\Delta\psi$.

The Mercury's perihelion ''anomalous'' shift was found out as a result of many-year diligent work of French astronomer \textit{Urbain Jean Joseph Le Verrier} who observed the Mercury at the Paris Observatory during almost half a century and elaborated the theory of its motion taking into account all the gravitational perturbations he knew at that time~\cite{1859:article:LeVerrier}. Le Verrier revealed that the calculated Mercury's trajectory was somewhat different from the optical observations, namely, the perihelion was being displaced faster than predicted by his theory. The obtained discrepancy appeared to be~$\Delta\psi\!\approx0.1''$ for the Mercury synodic period. To facilitate comparison of this small quantity in analyzing different hypotheses explaining the nature of discrepancy~$\Delta\psi$, it was decided to express it in terms of time periods equal to $100$ Earth years. Different authors report the discrepancy values ranging from $38''$ to $43''$.

Unfortunately, the classical mechanics has ''almost officially'' refused further investigation of the~$\Delta\psi$ nature by accepting, without proper criticism, in the early 20th century, the Einstein's formula~\cite{1915:article:EinsteinErklrungDP,%
	1938:book:Einstein:Infeld,%
	1921:book:Pauli:de}
for the ''anomalous'' Mercury's perihelion shift looking like
\begin{equation}\label{eq:apsides7:dpsi:Einstein}
	\Delta\psi=\frac{6\pi\G M_\Sun}{c^2 a (1-e^2)}
	\qquad\text{or}\qquad
	\Delta\psi=\frac{24\pi^3a^2}{c^2 T^2 (1-e^2)}\:,
\end{equation}
where $\G$ is the gravitational constant, $M_\Sun$ is the mass of the Sun as a material point, $c$ is the light velocity in vacuum, $T$ is the Mercury's synodic period, $a$, $e$ are the semi-major axis and eccentricity of the unperturbed Mercury's orbit, respectively.

Relationship~\eqref{eq:apsides7:dpsi:Einstein} suggested by Einstein gave the required value of the Mercury's perihelion shift~$\Delta\psi\approx 0.1''$ , however, at the cost of refusing the classical Newton's law on gravitational interaction. At that, any search for gravitation anomalies or other physical processes able to explain the~$\Delta\psi$ nature was finished. Formula~\eqref{eq:apsides7:dpsi:Einstein} was accepted by the scientific community as the only possible explanation of the Le~Verrier's discrepancy between the calculated and actual locations of the Mercury's perihelion equal to\\ $\Delta\psi\approx 0.1''$ per one revolution about the Sun.

Analysis of the postulates and hypotheses formula~\eqref{eq:apsides7:dpsi:Einstein} for constant~$\Delta\psi$ is based on is not the objective of this study. All the history of search for sources of the Mercury's perihelion ''anomalous'' shift is given in detail in the book by N.~T.~Roseveare ''\textit{Mercury’s perihelion: from Le Verrier to Einstein}''~\cite{1982:Roseveare:history:en}.

To our mind, the classical mechanics made haste to give up its position. The Le~Verrier's theory and Einstein's formula~\eqref{eq:apsides7:dpsi:Einstein} consider the Sun as a material point. But are there sufficient grounds to assume the Sun to be a material point in calculating the Mercury's orbit? The minimal distance between the Mercury and Sun is about~$66$ Sun radii or $153$~light seconds, which is very close to the Sun on the cosmic scale. Thus it is possible to regard the Sun not as a material point but as a deformable body whose shape may differ from a sphere.

Thus, we have a closed system of two gravitating bodies: the Sun and Mercury. Estimates obtained based on observations show that the Sun's self-rotation%
\footnote{%
	The self-rotation period of the Sun or, more exactly, of  its surface layers,\\ is $25$ to $34$~days.
}
causes its polar compression, which means that the Sun's polar radius is shorter than the equatorial one by $\approx6.3$~\textit{km}. In the case of an ellipsoid of revolution, including the case of the Sun, the compression estimate will be introduced via the oblateness%
\footnote{
	This a quantity characterizing the extent of deformation of a circle or sphere diameter with formation of an ellipse or ellipsoid. The oblateness is typically designed as~$f$. It is defined as $f=\frac{a-b}{a}$ where $a$ and $b$ are the ellipse semi-major and  semi-minor axes. Deformation (oblateness) of planets depends on their structures and self-rotation velocities.
}.
The existence of the compression/oblateness means that the assumption of the Sun deformability is well-grounded. Now, the Sun is a deformable body while the Mercury is a material point (proofmass).

As a component of the Solar System, the Sun participates in two motions, namely, its self-rotation and revolution about the barycenter%
\footnote{
	Instantaneous Solar System center of mass.
}.
The effect of the Sun bulges caused by its self-rotation upon the Mercury's orbit is similar to that of the Earth's bulges upon the trajectories of its artificial satellites~\cite{1980:book:Levantovsky:en,1977:book:Aksenov:ISZ:en}. However, since the Sun's oblateness~($\nicefrac{1}{9000000}$) is significantly (approximately $30200$ times) lower than that of the Earth~($\nicefrac{1}{298}$), the gravitational effect of the Sun's bulges on the Mercury's orbit may be ignored. A~similar assumption was made in~\cite[Chapter~7]{1976:book:M.G.Bowler:gravitation}.

\begin{figure}[!htb]
	\centering
	\includegraphics[scale=0.82]{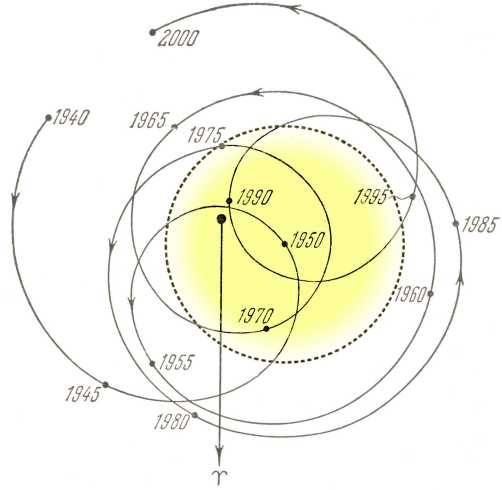}
	\caption{%
		The picture has been taken from book~\cite{1953:book:Kulikovsky:en}. ''\textit{Fig.~22. The Sun's center motion relative to the Solar System center of mass from $1940$ to $2000$.}''
	}
	\label{fig:apsides7-fig-scan-CMSS-1953-Kulikovsky}
\end{figure}

Thus, the only remaining type of rotation is revolution of the Sun as a deformable body about the Solar System barycenter whose location continuously changes with respect to the Sun's center of mass (Fig.~\ref{fig:apsides7-fig-scan-CMSS-1953-Kulikovsky}).

Like all the Solar System material objects, the Sun is affected by centrifugal forces in rotating about the Solar System barycenter. Therefore, it stands to reason that the Sun as a deformable body obtains under the action of centrifugal forces a shape different from the sphere. In the first approximation, this shape may be assumed to be an ellipsoid whose semi-major axis lies in the ecliptic plane and is directed to the Solar System barycenter.

\begin{figure}[h!]
	\centering\includegraphics[scale=1.2]{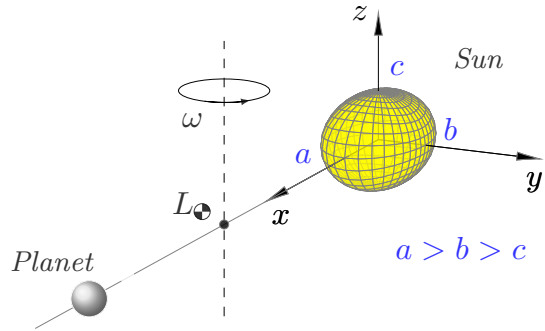}
	\caption{%
		Rotation of the Sun-Planet System about center of mass~$L_{\centerofmass}$.
	}
	\label{fig:apsides7-fig-barycenter}
\end{figure}

Let us estimate the effect of each planet on the center of mass~$L_{\centerofmass}$ position relative to the Sun (Fig.~\ref{fig:apsides7-fig-barycenter}).
The~$L_{\centerofmass}$ calculations for each planet%
\footnote{%
	The planet characteristics have been taken from \url{http://nssdc.gsfc.nasa.gov}
}
are given in Table~\ref{tab:apsides7:SSParameters}.

\begin{table}[!ht]\label{tab:apsides7:SSParameters}
	\centering
	\footnotesize
	\topcaption{%
		The ratio between the center of mass~$L_{\centerofmass}$ shift and average Sun radius~$R_\Sun$ for each Solar System planet. $L$~is the distance between the Sun and planet in light minutes, $e$~is the orbit eccentricity, $T$~is the planet synodic period.
	}
	\begin{tabular}{l|rrcccrcc}
		\rule{0pt}{3ex}
		&& $L$, [$^\ast$\textit{min}] && $e$ && $T$, [\textit{day}] && $L_{\centerofmass} / R_\Sun$
		\rule[-1.5ex]{0pt}{0pt}\\
	 	\hline
	 	\rule{0pt}{4ex}%
		Mercury && $  3.219$ && $0.20566$ && $   87.969$ && $0.00001$\\
		Venus   && $  6.016$ && $0.00675$ && $  224.701$ && $0.00038$ \\
		Earth    && $  8.317$ && $0.01675$ && $  365.256$ && $0.00065$ \\
		Mars     && $ 12.671$ && $0.09347$ && $  686.980$ && $0.00011$ \\
\textbf{Jupiter}  && $ 43.284$ && $0.04887$ && $\mathbf{4332.589}$ && $\mathbf{1.06733}$ \\
		Saturn   && $ 79.695$ && $0.05649$ && $10759.220$ && $0.58876$ \\
		Uranus     && $159.692$ && $0.04566$ && $30685.400$ && $0.18025$ \\
		Neptune   && $249.898$ && $0.01126$ && $60189.000$ && $0.33275$ \\
		Pluto   && $328.359$ && $0.24881$ && $90560.000$ && $0.00006$\rule[-2.0ex]{0pt}{0pt}\\
	\end{tabular}
\end{table}

Table~\ref{tab:apsides7:SSParameters} demonstrates that the main factor determining the radial displacement of the Solar System barycenter relative to the Sun's center of mass are the giant planets, namely, \textit{Jupiter}, \textit{Saturn}, \textit{Uranus} and \textit{Neptune}. Certainly, allowing for the current giant planet locations in determining the barycenter coordinates relative to the Sun will introduce some corrections. However, the very fact of the existence of the variable field of centrifugal forces acting upon the Sun in the planes parallel to the ecliptic plane cannot be ignored. Since among all the planets only Jupiter produces the maximal displacement of center of mass~$L_{\centerofmass}$ with respect to the Sun, let us consider Jupiter as the main source of the Sun's centrifugal deformation that, as we believe, will finally allow revealing the nature of the Mercury's perihelion ''anomalous'' shift~$\Delta\psi=0.1''$.

The period of the Sun--Jupiter pair revolution about~$L_{\centerofmass}$ is determined by Jupiter's synodic period. Hence, in rotating about the center of mass of the Sun--Jupiter system, the Sun as a deformable body will be affected by relevant centrifugal forces that form, in the first approximation, a three-axis ellipsoid%
\footnote{%
	In reality, the ''liquid'' Sun shape is much more complicated~\cite{1903:book:Poincare:figures, 1950:book:Krat:en} since there is also the self-rotation as well as continuously varying position of the Solar System barycenter. However, this is beyond the scope of our study.
}
with semi-major axis~$a$ lying in the ecliptic plane and oriented on Jupiter.

Actually, we consider a three-body system: Sun, Mercury and Jupiter, but in two stages. At the first stage we justify the Sun's nonsphericity by the existence of centrifugal forces when the Sun--Jupiter pair revolves about the center of mass with the period equal to the Jupiter's synodic period. At the second stage we analyze the Mercury's motion about the center of mass of the ellipsoid of Sun. The self-rotation periods of the Sun--Jupiter and Sun--Mercury pairs are quite different. During one Mercury's revolution about the Sun, the Sun itself moves with respect to center of mass~$L_{\centerofmass}$ of the Sun--Jupiter pair by the angle equal to
\begin{equation}\label{eq:apsides7:estimation}
	\frac{2\pi}{T_\Jupiter}\; T_\Mercury \cdot
	\left(\frac{180}{\pi}\right)\approx 7.3^\degree\:,
\end{equation}
where $T_\Jupiter$ and $T_\Mercury$ are the synodic periods of Jupiter and Mercury. The angle is quite small. In the first approximation we can assume that the Sun is a motionless ellipsoid from the Mercury's point of view.

Thus, the subject of research is a closed system consisting of the Sun and Mercury. Mercury as a material point revolves about the motionless ellipsoid of the Sun. Our goal is to reveal the possible relationship between the Sun's ellipsoid parameters and ''anomalous'' shift of Mercury's perihelion~$\Delta\psi$ provided such relationship indeed exists.


%

\section{A system of two gravitating bodies}

\begin{figure}[htb]
	\centering
	\includegraphics[scale=1]{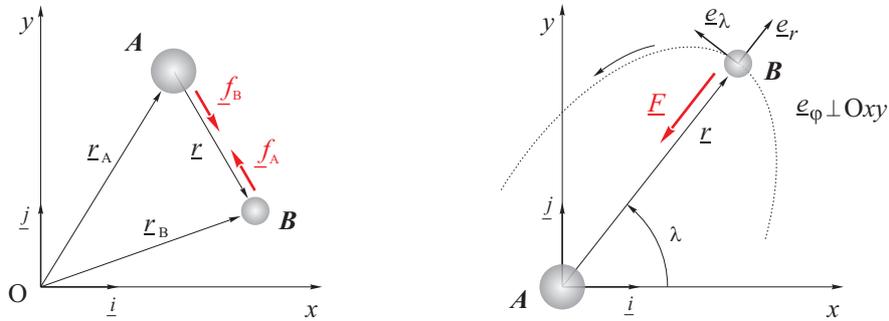}
	\caption{%
		The problem of unperturbed motion of two gravitating bodies $A$ and $B$ in plane~$\mathbf{O}xy$. In the left panel, $A$ and $B$ move with respect to point $\mathbf{O}$, while in the right panel $A$ is motionless and its center of mass coincides with the coordinate system origin $\mathbf{O}$. $\vec{i},\vec{j},\vec{k}$ are unit vectors of the Cartesian frame of reference $\mathbf{O}xyz$, $\vec{e}_r,\vec{e}_\varphi,\vec{e}_\lambda$ are unit vectors of the spherical coordinate system where the body $B$ position is defined by coordinates $r,\varphi,\lambda$. Under the condition of our task~$\varphi\equiv 0$.
	}
	\label{fig:apsides7-fig-mAmB}
\end{figure}

\noindent
Consider a classical problem of motion of two gravitating bodies obeying the Newton's mechanics laws. Let us inject into consideration a motionless orthogonal coordinate system~$\mathbf{O}xyz$ (Fig.~\ref{fig:apsides7-fig-mAmB}). Radius-vectors $\vec{r}_A$ and $\vec{r}_B$ define the current positions of bodies $A$ and $B$ with masses $m_A$ and $m_B$, respectively. Let bodies $A$ and $B$ be centrally symmetric.

Behavior of gravitationally interacting bodies $A$ and $B$ in the Fig.~\ref{fig:apsides7-fig-mAmB} left panel may be described by a set of second-order differential equations
\begin{equation}\label{eq:apsides7:system-mAmB-left}
	\left\{
	\begin{array}{lcl}
		m_A\:\vec{\ddot{r}}_A &=& \vec{f}_{B} \\
		m_B\:\vec{\ddot{r}}_B &=& \vec{f}_{A}
	\end{array}\right.
	\quad\text{or}\qquad
	\left\{
	\begin{array}{lcl}
		m_A\:\vec{\ddot{r}}_A &=& \vec{E}_{B}\:m_A \\
		m_B\:\vec{\ddot{r}}_B &=& \vec{E}_{A}\:m_B
	\end{array}\right.\qquad
\end{equation}
Here $\vec{f}_{B}$ is the gravitational force acting on body $A$ from body~$B$, which is a product of the gravity field intensity~$\vec{E}_B$ and mass~$m_A$. Force~$\vec{f}_{A}$ is defined in the same way.

Let us simplify task~\eqref{eq:apsides7:system-mAmB-left} by subtracting the upper equation from the lower one. Taking into account that
\begin{equation}\label{eq:apsides7:r}
	\vec{r}=\vec{r}_B-\vec{r}_A
\end{equation}
obtain the equation of the body~$B$ motion with respect to the motionless centrally-symmetric body~$A$
\begin{equation}\label{eq:apsides7:system-mAmB-right-EAEB}
	\vec{\ddot{r}} = \vec{E}_A - \vec{E}_B\;.
\end{equation}

Gravity field intensities $\vec{E}_A$ and $\vec{E}_B$ will be expressed through corresponding potential gradients~$\mathbf{U}_A$ and $\mathbf{U}_B$
\begin{equation}\label{eq:apsides7:AEU}
	\mathbf{U}_A=\G\frac{\:m_A\:}{r}\:,\;\;\;
	\vec{E}_A=\bigl(\nabla \mathbf{U}_A \cdot \vec{e}_r \bigr)\; \vec{e}_r
	\quad
	\Rightarrow
	\quad
	\vec{E}_A=-\frac{\:\G m_A\:}{r^2}\; \vec{e}_r\:.
\end{equation}
\begin{equation}\label{eq:apsides7:BEU}
	\mathbf{U}_B=\G\frac{\:m_B\:}{r}\:,\;\;\;
	\vec{E}_B=\bigl(\nabla \mathbf{U}_B \cdot \vec{e}_r\bigr)(-\vec{e}_r)
	\quad
	\Rightarrow
	\quad
	\vec{E}_B=-\frac{\:\G m_B\:}{r^2}\left(-\vec{e}_r\right)\:.
\end{equation}
Here $\G$ is the gravitational constant, $r=|\vec{r}|$ is the distance between point masses $m_A$ and $m_B$, $\nabla$ is the nabla operator that is a vector differential operator in the spherical coordinate system
\begin{equation}\label{eq:apsides7:nabla}
	\nabla =
	\frac{\partial}{\:\partial r\:}\vec{e}_r +
	\frac{1}{r}\frac{\partial}{\:\partial \varphi\:}\vec{e}_\varphi +
	\frac{1}{\:r\cos\varphi\:}\frac{\partial}{\:\partial \lambda\:}\vec{e}_\lambda\;,
\end{equation}
where $\vec{e}_r,\vec{e}_\varphi,\vec{e}_\lambda$ are unit vectors of the spherical coordinate system (Fig.~\ref{fig:apsides7-fig-mAmB}).

Let us express relations \eqref{eq:apsides7:AEU} and \eqref{eq:apsides7:BEU} obtained for the bodies $A$ and $B$ gravity field in terms of potentials and substitute them into equation~\eqref{eq:apsides7:system-mAmB-right-EAEB}:
\begin{equation}\label{eq:apsides7:system-mAmB-right-UAUB}
	\vec{\ddot{r}} =
		\bigl(\nabla \left(\mathbf{U}_A+\mathbf{U}_B\right)\cdot\vec{e}_r\bigr)\:\vec{e}_r
\end{equation}

Substituting \eqref{eq:apsides7:AEU} and \eqref{eq:apsides7:BEU} into~\eqref{eq:apsides7:system-mAmB-right-EAEB} or \eqref{eq:apsides7:system-mAmB-right-UAUB} and taking into account the fact that $m_A \gg m_B$ under the task condition, obtain
\begin{equation}\label{eq:apsides7:system-mAmB-right}
	\vec{\ddot{r}} =
	-\G\:\frac{\:m_A+m_B\:}{r^2}\:\vec{e}_r
	\quad
	\Longrightarrow
	\quad
	\vec{\ddot{r}} \approx
	-\G\:\frac{\:m_A\:}{r^2}\:\vec{e}_r
\end{equation}
Thus, excluding from~\eqref{eq:apsides7:system-mAmB-right} mass $m_B$, we exclude from consideration the body~$B$ gravity field itself.

\begin{figure}[h]
	\centering
	\includegraphics[scale=1.1]{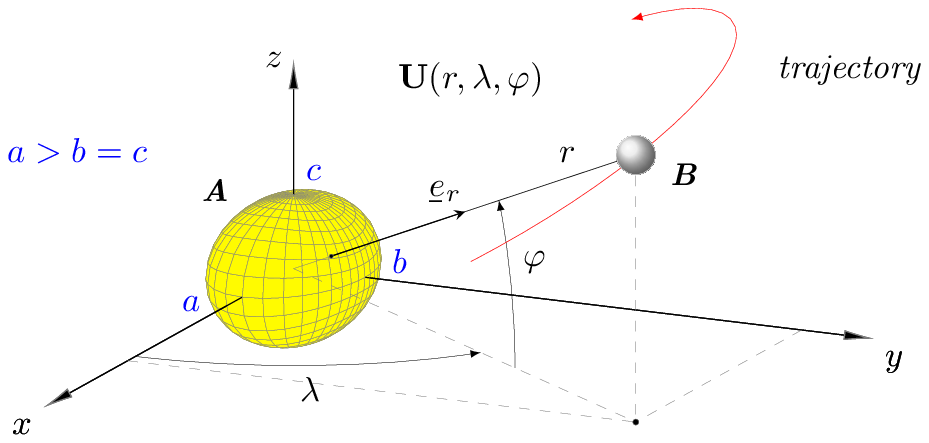}
	\caption{}
	\label{fig:apsides7-fig-ellipsoid}
\end{figure}

However, obtained equation \eqref{eq:apsides7:system-mAmB-right} is valid only for point masses or centrally-symmetric bodies. To factor in the influence of the central body~$A$ (mass~$m_A$) geometry on the trajectory of point mass~$m_B$, let us use in the first approximation the expression for the three-axis ellipsoid potential. In general case, the three-axis ellipsoid potential is a convergent series expressed in terms of Legendre polynomials~\cite{1983:book:Grushinsky:en,1989:book:Torg}. For convenience  and without compromising the essence, let us orient the ellipsoid so that the axes of the Cartesian coordinate system~$\mathbf{O}xyz$ coincide with the  ellipsoid's  principal axes, namely, the ellipsoid semi-axis $a$ coincides with axis~$\mathbf{O}x$, semi-axis $b$ coincides with axis~$\mathbf{O}y$, and semi-axis $c$ lies on axis~$\mathbf{O}z$. The coordinate system origin (point~$\mathbf{O}$) coincides with the center of mass of the homogeneous ellipsoid (Fig.~\ref{fig:apsides7-fig-ellipsoid}).
Let us restrict ourselves to three expansion terms of the three-axis ellipsoid potential. This is sufficient to allow for the shape of central body~$A$ in calculating the satellite~$B$ trajectory, i.e.,
\begin{equation}\label{eq:apsides7:Uabc}
	\mathbf{U}=\frac{\G m_A}{r}+
		\frac{\G}{r^3}\:
		\Bigg(\!\!
			\left(J_c-\frac{J_a\!+\!J_b}{2}\right)\frac{1-\sin^2\varphi}{2}+
			\left(J_b\!-\!J_a\right)\frac{3cos^2\varphi\cos 2\lambda}{4}
		\Bigg)\:,
\end{equation}
where
\begin{equation}\label{eq:apsides7:Jabc}
	J_a=\frac{m_A}{5}\left(b^2+c^2\right)\:,\quad
	J_b=\frac{m_A}{5}\left(a^2+c^2\right)\:,\quad
	J_c=\frac{m_A}{5}\left(a^2+b^2\right)\:.
\end{equation}
Here $J_a$, $J_b$, $J_c$ are the ellipsoid's principal moments of inertia,   $\G$ is the scale-dimension factor (gravitational constant), $m_A$ is the ellipsoid's gravitating mass, $r,\varphi,\lambda$ are the spherical coordinates of material point~$B$, i.e., distance, latitude and longitude, respectively.

Let us make some assumptions that will help simplify expression~\eqref{eq:apsides7:Uabc} for the potential of central body~$A$ which is a three-axis ellipsoid
\begin{enumerate}\itemsep -3pt
\item
	$\varphi(t)\equiv 0\;$ (trajectory $B$ lies in plane $\mathbf{O}xy$)
\item
	$a>b=c\;$ (ellipsoid of revolution)
\end{enumerate}

As a result, expression~\eqref{eq:apsides7:Uabc} takes a simpler form:
\begin{equation}\label{eq:apsides7:Uabc-Oxy}
\boxed{\quad
	\mathbf{U}(r,\lambda)=\G\:\frac{m_A}{r}
	\Biggl(
	1+\frac{1+3\cos 2\lambda}{r^2}\left(\frac{a^2-b^2}{20}\right)
	\Biggr)\quad}
\end{equation}

Thus, in our task the equation~\eqref{eq:apsides7:system-mAmB-right-UAUB} potentials will be as follows:
\begin{equation}\label{eq:apsides7:UAUB}
	\mathbf{U}_A=\mathbf{U}(r,\lambda)\;,\quad
	\mathbf{U}_B=0\;.
\end{equation}
Substituting~\eqref{eq:apsides7:UAUB} into~\eqref{eq:apsides7:system-mAmB-right-UAUB}, obtain the motion equation for material point~$B$ having radius-vector~$\vec{r}$ with respect to motionless central body~$A$ shaped as an ellipsoid with semi-axes $a>b\!=\!c$:
\begin{equation}\label{eq:apsides7:F}
\boxed{
	\quad
	\vec{\ddot{r}}=-\;\G\:\frac{m_A}{r^2}
	\Biggl(
		1+3\left(\frac{a^2-b^2}{20}\right)
		\frac{1+3\cos 2\lambda}{r^2}
	\Biggr)\vec{e}_r\;.
}
\end{equation}

The ellipsoid deformation is defined by changes in its semi-major axis~$a$. Semi-axes $b$ and $c$ will be recalculated under the condition of invariance of the ellipsoid volume~$V_A$ in the process of its deformation.
Hence, when semi-major axis $a$ is varied, semi-axes $b$ and $c$ are to vary correspondingly:
\begin{equation}\label{eq:apsides7:Volume}
	V_A(k)=\frac{4}{3}\pi abc=\mathit{Const}
	\quad\implies\quad
	a=k R_A\:,\;\;
	b=\frac{R_A}{\sqrt{k\:}}\:,\;
	c=\frac{R_A}{\sqrt{k\:}}\:.
\end{equation}
Here $R_A$ is the radius of the undeformed spherical body~$A$, $k\!>\!1$ is the real factor defining the extent of the ellipsoid transformation via variation in the length of its semi-axis~$a$.


%

\section{Research results}

\noindent
In this study, we have solved numerically equation~\eqref{eq:apsides7:F} where central body~$А$ is assumed to be the Sun and material point~$B$ represents Mercury. The initial conditions for Mercury are as follows:
\begin{equation}\label{eq:apsides7:system-mAmB-right-ic}
	\lambda(t) \Big|_{t=0}\! = \lambda_0\;,\quad
	r(t)\Big|_{t=0}       \! = r_\text{perihelion}\:,\quad
	\dot{r}(t) \Big|_{t=0}\! = v_\text{perihelion}\;.
\end{equation}

The procedure for solving equation \eqref{eq:apsides7:F} with initial conditions~\eqref{eq:apsides7:system-mAmB-right-ic} is a sequence of calculating the Mercury's trajectory at different values of semi-axis $a$ of the ellipsoid, that is, the Sun.
$$
	a(k)=k\:R_\Sun\;,\quad\text{where}\quad k\in\mathbb{R}\;,\;\;\; k>1\;.
$$
Among a great variety of solutions at $\lambda_0=0$, the following value of the deformation coefficient $k$ was chosen:
\begin{equation}\label{eq:apsides7:system-mAmB-right-k}
	\boxed{\quad k\approx\!1.00044 \quad}
\end{equation}
at which the so-called ''anomalous'' shift of the Mercury perihelion~$\Delta\psi=0.1''$ takes place. This means that the Sun's oblateness in the ecliptic plane is about~$\nicefrac{1}{1516}$, namely, semi-major axis $a$~exceeds semi-axis $b$ by $\approx\!459$\textit{~km}. As compared with the Sun size, this is a quite small value, $\approx\!0.07\%$ of the Sun's radius. However, just this difference in the Sun's ellipsoid semi-axes gives rise to the missing Mercury's anomaly.

\begin{figure}[h]
	\centering
	\includegraphics[scale=1]{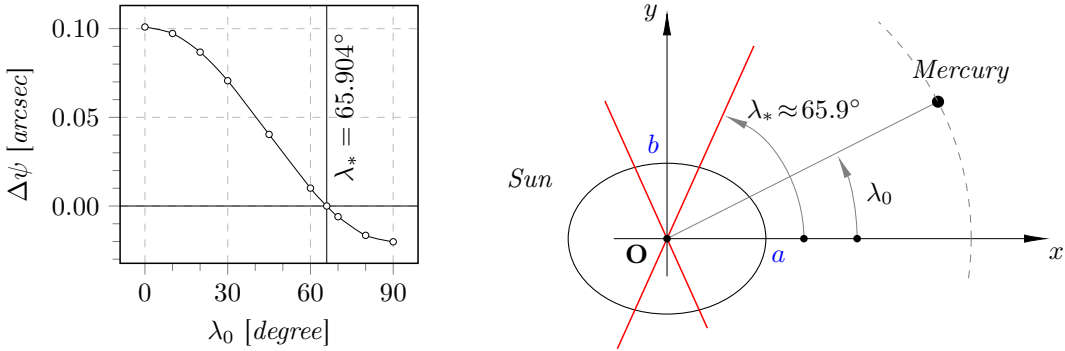}
	\caption{%
		The effect of the angle $\lambda_0$ value on the Mercury's perihelion shift~$\Delta\psi$. $\lambda\ast$ is the critical angle at which the perihelion shift direction changes.
	}
	\label{fig:apsides7-fig-orbit-lambda-dpsi}
\end{figure}

In addition, in solving the problem we have revealed a dependence of the Mercury's perihelion direction on the initial angle~$\lambda_0$ in~\eqref{eq:apsides7:system-mAmB-right-ic}. This result is shown graphically in Fig.~\ref{fig:apsides7-fig-orbit-lambda-dpsi}.
The plot shows that when $\lambda_0$ belongs to ranges
\begin{equation}\label{eq:apsides7:system-mAmB-lambda0}
	\lambda_\ast < \lambda_0 < \pi\!-\!\lambda_\ast
	\qquad\text{and}\qquad
	\pi\!+\!\lambda_\ast < \lambda_0 < 2\pi\!-\!\lambda_\ast\;,
\end{equation}
the reverse (clockwise) motion of the Mercury's perihelion is observed only conceptually.


%

\section{Conclusions}

\noindent
The performed study of the effect of the Sun's nonsphericity on the Mercury's trajectory has shown that the so-called ''anomalous'' shift of the Mercury's perihelion, that is, the angular discrepancy, is a result of the fact that Le~Verrier assumed in his calculations that the Sun is a material point.

The Mercury's perihelion shift~$\Delta\psi\approx 0.1''$ is determined by the Sun's nonsphericity characterized by the ratio between the ellipsoid semi-axes $a$,$b$ and~$c$. The source of the Sun's nonsphericity (when this is not about the polar compression) is the Sun deformation in its rotation about the Solar System barycenter.

The practice of satellite-based gravity prospecting of the Earth, Moon and Mars shows that the artificial satellite trajectory is quite sensitive to gravity anomalies of the central body. In our case, the role of the satellite indicating the Sun's gravity anomalies is played by Mercury. The real indicator of the Sun's gravity anomaly is the Mercury's angular discrepancy~$\Delta\psi\approx 0.1''$.

The fact that Mercury in its motion is sensitive to the Sun's gravity anomalies makes it clear to us that, if peculiar features of the Sun deformation are taken into account in the theoretical model of Mercury's motion, then the ''anomalous'' shift detected by \textit{Urbain Le~Verrier} merely disappears. Note that this was revealed without involving metaphysics or modifying the Newton's law on the gravitational interaction of material point.

Actually, the problem of interpretation of ''anomalous'' shift~$\Delta\psi$  reduces to the inverse satellite-based gravity prospecting problem~\cite{1980:book:Mironov:en} as applied to the Sun and its satellite Mercury.

\section*{Afterword}

\noindent
In solving this problem with a ''great'' history, as well as in solving other problems, we followed the fact that in a case of an unsolvable, at first glance, problem, one should not hurry to involve metaphysics and construct something abstract and immaterial. A Scientist should in any situation remain on the materialistic position and base himself on facts confirmed by independent experiments. May be it looks routinely, but it is indeed necessary to be critical on the existing metaphysic interpretations of observed phenomena, which further become a basis for one or another metaphysical concept, e.g., ''dark matter''.

If the problem remains unsolved for decades or even centuries, then one should return to the initial point, to the beginning of the way, and try to find new or previously missed facts, conduct test experiments and observations, and check the correctness of the problem definition.
\begin{quotation}
	\textit{If you desire to return to the truth you do not need to search for the road. You know it. You came down that way. Retrace your footsteps.}\par
	\hfill \textsc{\footnotesize Bernard of Clairvaux}
\end{quotation}

Diligently following this principle, we succeeded, as we hope, in solving several problems with history, namely, in revealing the sources of Chandler's wobble of the Earth's rotation pole~\cite{2014:PAMM:PAMM201410017}, in answering the question of why nobody succeeded in refining the gravitation constant~\cite{2017:viXra:1708.0245v3:GConst}, in proving the real existence of the gravitational dipole and formulating the definition of the gravitating mass~\cite{2014:viXra:1406.0128v2:MASS7:english}, in explaining physical nature of the ''plateau'' in the galaxy rotation curve without using such a metaphysical concept as ''dark matter''~\cite{2018arXiv181106372K}.


\begin{spacing}{0.90}
	\bibliographystyle{ieeetr} 
	\bibliography{%
		../../../../../___General-LATEX-BIB/Kiryan-bibliography-prive-utf8,
		../../../../../___General-LATEX-BIB/Kiryan-bibliography-general-utf8}
\end{spacing}

\end{document}